# Field-effect and frequency dependent transport in semiconductor-enriched single-wall carbon nanotube network device

Manu Jaiswal[1,§], C.S. Suchand Sangeeth[1,*], Wei Wang[2], Ya-Ping Sun[2] and Reghu Menon[1]

[1] Department of Physics, Indian Institute of Science, Bangalore 560012, India

[2] Department of Chemistry and Laboratory for Emerging Materials and Technology, Clemson University, Clemson SC 29634-0973, USA

[§] Present address: Max Planck Institute for Polymer Research, Mainz D55021, Germany

Email: suchand@physics.iisc.ernet.in

## Abstract

The electrical and optical response of a field-effect device comprising a network of semiconductor-enriched single-wall carbon nanotubes, gated with sodium chloride solution is investigated. Field-effect is demonstrated in a device that uses facile fabrication techniques along with a small-ion as the gate electrolyte – and this is accomplished as a result of the semiconductor enhancement of the tubes. The optical transparency and electrical resistance of the device are modulated with gate voltage. A time-response study of the modulation of optical transparency and electrical resistance upon application of gate voltage suggests the percolative charge transport in the network. Also the ac response in the network is investigated as a function of frequency and temperature down to 5 K. An empirical relation between onset frequency and temperature is determined.

## 1. INTRODUCTION

Single-walled carbon nanotubes (SWNT) have a significant potential for applications in electronic devices since their unique morphological characteristics and aspect ratio lead to high mobility, transconductance, and good device characteristics. Additionally, their performance shows sensitivity to a variety of materials from organic gases to ionic solution – leading to possible sensor applications. These device applications are however not without limitations. Field-effect devices based on individual nanotubes work well but are constrained by difficult process involved in contacting the individual tube. In case of a network of tubes, the various fabrication procedures usually lead to bundling of the tubes. The presence of metallic tubes (in pristine SWNTs, typically 33 % are metallic) within these bundles subdues the field-effect on the semiconducting tubes.[1, 2] The field-effect is therefore very small and also dependent on detailed fabrication processes, designed to reduce the extent of inhomogeneity and bundling. Whereas percolation governs the transport in the network, a detailed low temperature investigation is needed to understand this phenomenon.

Electrochemical gating of FETs has been used on polymeric and nanotube systems.[3-5] Nanotubes can show amphoteric doping behavior. The shift of Fermi level with hole/electron doping with an electrolyte was first reported on a single rope of multi-wall carbon nanotubes (MWNTs). It was found that better gating could be achieved with liquid-ion gating, as compared to a solid gate dielectric.[4] The former allows contact with a greater volume of tubes despite the slow response time. The ion molecules did not chemically bind to the nanotubes and the conductivity is recoverable after the solution is removed. It has been observed that SWNT films undergo charge transfer and the experimental results have been interpreted by invoking a simple electron transfer mechanism within a rigid-band model.[6] A single tube of SWNT tube can be electrochemically gated with NaCl solution[7, 8] and an optical analog of the transparent SWNT FET has also be made with a large-ionic liquid gate electrolyte and using vacuum-filtered film deposition.[9]

The transparency modulation with gating in a semiconductor-enriched SWNT network is studied in detail in this paper. It is useful to enhance the fraction of semiconducting tubes since the van Hove singularity transitions of semiconducting tubes are significant in the optical transparency modulation. The residual metallic tubes in the network form conducting pathways which reduce the field-effect. Using this enhancement procedure, a field-effect device is fabricated that utilizes facile fabrication techniques (by air-spraying of tubes on to a heated glass slide) together with a commonly available lab reagent as the gate electrolyte (a small ion solution operating at low voltages). A comparative investigation of the electrical and optical time response is made for such a device. Further, this paper studies the role of temperature in the percolation phenomenon in the network and this is done with an ac impedance investigation.

## 2. EXPERIMENTAL DETAILS

The SWNTs are prepared by carbon-arc method and comprise of metallic and semiconducting nanotubes in the ratio 1:2. The fraction of semiconducting nanotubes is enhanced by means of separating agents that preferentially associate with them on account of the static charge present on semiconducting tubes. This yields nearly ~ 90% enriched semiconducting SWNTs, and the process is described in detail elsewhere.[10-12]

The nanotube network was prepared by air-spraying a dispersion of the enriched SWNTs in dimethylformamide (DMF) onto a heated clean glass slide. The nanotubes are randomly aligned in bundles of diameter 10-20 nm to form a thin interconnected network with high transparency (> 80 %). [12, 13]

The field-effect device is constructed according to a design originally suggested by Wu et al.[9] and it involves the use of an electrolyte gate in contact with a remote electrode. A small-ion electrolyte (0.1 M NaCl solution) however is used for gating this nanotube FET. Individual nanotube FETs have been shown to operate in salty environments[7] and charged molecules can act as an effective gate that changes the conductance of the nanotubes. For the construction of a device, a rectangular piece of the transparent nanotube sheet on glass substrate is first separated into two electrically-insulated halves. A Teflon gasket with a glass plate pasted on top forms a reservoir which includes portions of both these separated networks. After being positioned vertically, this reservoir is filled with the ionic solution (~ 1 ml). Electrodes serving as gate and source are put on the other end of each of these separated halves. An additional 'drain' electrode is fabricated close to the source electrode to measure the resistance change in the nanotube networks due to gating. Application of a gate voltage draws ions from the NaCl solution towards the nanotubes at the source / drain terminals by means of the formation of an electric-double layer, and the ions of this double layer behave as an exceptionally near-lying gate electrode. Only small gate voltages ($V_g < 0.83$ volts) can be applied for the device to prevent redox reactions from occurring in the gate electrolyte solution.

The transparency of the device is measured as function of wavelength in the visible and near-IR spectral regions using a Bruker IFS 66 v/S optical spectrophotometer. To modulate the transmittance, external bias voltages are applied between the source and gate of the device using a Keithley-2400 Sourcemeter and the charging current is also monitored. The spectrometer records the variation in transparency in the wavelengths of interest, as a function of the applied gate voltages.

The time-dependent optical response of the device is investigated using the rapid scan time-resolved measurement option available with the spectrometer over a region of wavelength (6250 cm$^{-1}$ to 5000 cm$^{-1}$) - where transparency modulation is observed. The spectrometer sweeps this wavelength region at time intervals of 400 mS. A measurement of the transparency of the network in the region that is directly in contact with the NaCl reservoir is not possible due to large absorption by the electrolyte itself, so the measurement is performed in the adjacent region which is not directly in contact with the electrolyte.

The ac impedance of the nanotube network is measured with an Agilent 4285A LCR meter in the frequency range 100 KHz – 10 MHz, down to 5 K in a Janis continuous flow cryostat. The low frequency impedances (40 Hz-100 KHz) are measured using a dual channel lock-in amplifier (SR 830). A small ac signal (50 mV) is applied across the nanotube network through a 50 Ω resistor which is connected in series with the nanotube network. The lock-in amplifier is used to measure the voltage across the resistor and from this the current through the circuit is determined. The lock-in also records the in-phase voltage across the nanotube network, from which the real part of the complex impedance is estimated as a function of frequency.

## 3. RESULTS AND DISCUSSIONS

The electric-field between the nanotube at source / drain channel and the gate electrode shifts the nanotube Fermi level and modulates the charge carrier concentration. As a consequence, the transparency of the network is modulated by the application of gate voltage. The optical absorption spectrum of a SWNT film is characterized by two main features at approximately 1800 and 1000 nm, which are superimposed on the broad absorption band centered near 250 nm.[14] The density of states near Fermi level shows sharp features for carbon nanotubes, which correspond to the van Hove singularities. These optical features correspond to electronic transitions between pairs of van Hove singularities in the semiconducting SWNTs.

When a gate voltage is applied, charges are drawn from the source terminal onto the nanotube layer where oppositely charged ions line up in the solution at a distance given by the hydration radius of the ions. This double charge layer is associated with a capacitance. The FET charging (in a single MWNT) has been modeled by considering two capacitors connected in series.[4, 7] These correspond to two physical phenomena in the system – (a) External electric-field causing an electrostatic drop $\phi$ between the gate and the nanotube. (b) Shift in the chemical potential or Fermi energy $E_F$ (due to addition of charge) with respect to the charge neutrality point or middle-band gap energy ($E_F = 0$) – leading to a 'quantum capacitance' arising from the low density of states.

In Figure 1, the transmittance of the network is plotted as a function of applied gate voltages near the $S_{11}$ absorption region. Positive gate bias causes a progressive increase in transmittance around ~ 1800 nm as a result of reduction in hole concentration and consequent decrease in free carrier absorption. Negative voltage has the opposite effect of increasing the hole concentration (depletion of electrons from the van Hove singularity) in the film and causing a decrease in transmittance. The transmittance is modulated by ± 2 % for the highly transmitting films (transparency ~ 76 % at 1800 nm). It should be possible to obtain a larger modulation using an electrolyte that permits application of higher voltage or using a thicker sample with more material available to absorb the electromagnetic radiation at the expense of the reduction in transparency.

By attaching an additional contact near the source electrode, the resistance of the network is monitored as a function of voltage using standard two-probe method. The current is adjusted so that the measured voltage is much smaller than the voltages associated with gating. The resistance of the network is plotted as a function of gate bias in Figure 2. The resistance decreases (to half its value) when the Fermi level is gated away from the gap and increases in the opposite scenario. The resistance increase in the latter case is rather small (about 10 %) and the resistance tends to saturate. The positive gating depletes the semiconducting tubes thereby reducing their role in the electrical transport. However, the resistance change being small indicates that the semiconducting tubes contribute only a small fraction to the transport in the enriched-network at room temperature. This observation is also supported by magnetotransport studies on the SWNT network.[12, 13]

The network takes a time of the order of a few minutes to reach its final saturation value of absorption, for any applied gate voltage. The change in transparency with time for both directions of voltage ($V_g = \pm 600$ mV) is plotted in Figure 3. It can be seen that the transparency response is linear with time before tending to saturate. The rate of transparency change is different for the directions of gating – the transparency increases twice as faster than it decreases. This was further verified by applying a square wave of frequency ~ 0.1 to 1 Hz, which resulted in a net increase of transparency. The large time taken for the electrostatic equilibrium to be established can be attributed to time taken for charged ions to move in the solution and form a charged double layer at the interface with the tubes.

The time response for change in transparency and resistance for the network is compared in Figure 4. While the transparency changes slowly, there is faster response for the resistance change. The percolation pathways will readjust upon application of the gate voltage. Only a small fraction of tubes is involved in the formation of percolation pathway due to the high aspect ratio of the tubes. A lower-resistance percolation path may get established in a shorter time scale, whereas the transparency change represents a collective response of 'all' the tubes leading to a longer timescale.

To further investigate the role of percolation processes in the transparent SWNT networks, a study of the AC transport is also carried out. In Figure 5(a), the ac conductance is plotted at various temperatures down to 5 K. The data is normalized with the appropriate dc conductance value at that temperature. At low temperature, the ac conductance is a strong function of frequency. The extended pair approximation model is often used to describe the ac conductance in various disordered systems and the dependence of conductance on frequency in this model is given as:[15-18]

$$\sigma = \sigma_0[1 + k(\omega/\omega_0)^s] \tag{1}$$

This equation is valid close to the percolation threshold, when the system has a conducting mass fraction greater than the critical value at percolation. The data at various temperatures can only be approximately described within the extended pair approximation model as seen from the plot of master scaling curve in Figure 5(b). There is a significant deviation at 5 K, where the system may actually be very close to percolation threshold. As the temperature is decreased, increasing number of tubes are 'frozen out' from the conduction channels. The decrease in temperature can be considered as 'electrically equivalent' to the system coming closer to the percolation threshold. In other words, the system at lower temperature is electrically (but not optically) equivalent to a more transparent network scenario. The onset or critical frequency, $\omega_0$, can be defined as $\sigma(\omega_0) = 1.1\sigma_0$, where $\sigma_0$ is the dc conductivity and it is a measure of how close the system is to the percolation threshold. More transparent films have a lower value of onset frequency.[19] Xu et al. have shown a direct relationship between onset frequency and the dc conductivity,[19] and such a relationship also exists between the onset frequency and temperature as shown in the inset of Figure 5(b). Here, the log-log plot suggests a form, $\omega_0 \sim T^p$ for T < 100 K. A detailed investigation of percolation phenomenon in nanotube/polymer composite has been reported by Kilbride et al. using ac impedance measurements.[15] It is seen that the critical frequency $\omega_0$ decreases by many orders in magnitude as the network is rarefied. This shift in critical frequency to a lower value can be explained within the percolation mechanism. A carrier travels a distance $\xi$ at frequency $\omega_\xi$ and this length is reduced at higher frequencies. Above the critical frequency, some carriers can traverse the shorter distance within a well-connected region without requiring performing a difficult hop. A higher value of $\omega_0$ implies smaller correlation length and shorter connections in the system.[20] At lower temperatures, the connectivity in the system is reduced as shown by the shift in onset frequency to lower values. The AC transport can be a useful tool to probe 'aggregation' of tubes and bundles in the network, since both affect the onset frequency. This investigation in turn can be used for optimizing the fabrication process towards achieving a homogeneous network for FET applications, especially in sensors.

## 4. CONCLUSION

A facile construction of a SWNT network field-effect device and its operation with a commonly available small ion gate solution is demonstrated. The process relies on semiconductor enrichment of the nanotubes prior to their deposition on the substrate. The field-effect property is demonstrated using both optical and electrical effects. The temporal differences between the two phenomena, however, suggest the carrier transport in the system – wherein the latter alone depends on the percolation processes. The percolation processes are also investigated by a low temperature AC transport investigation of the network and suggest an approximate description within the extended pair model that shows deviation at the low temperatures (~ 5 K). An empirical relation between onset frequency and temperature is observed and the system at lower temperature is electrically equivalent to having a smaller density of tubes involved in effective transport.

**Acknowledgements:** The FTIR measurements were performed at the DST National facility, Bangalore. M.J. and C.S.S.S. thank CSIR, New Delhi for financial assistance. Y-PS acknowledges financial support from the US National Science Foundation.

**Figures:**

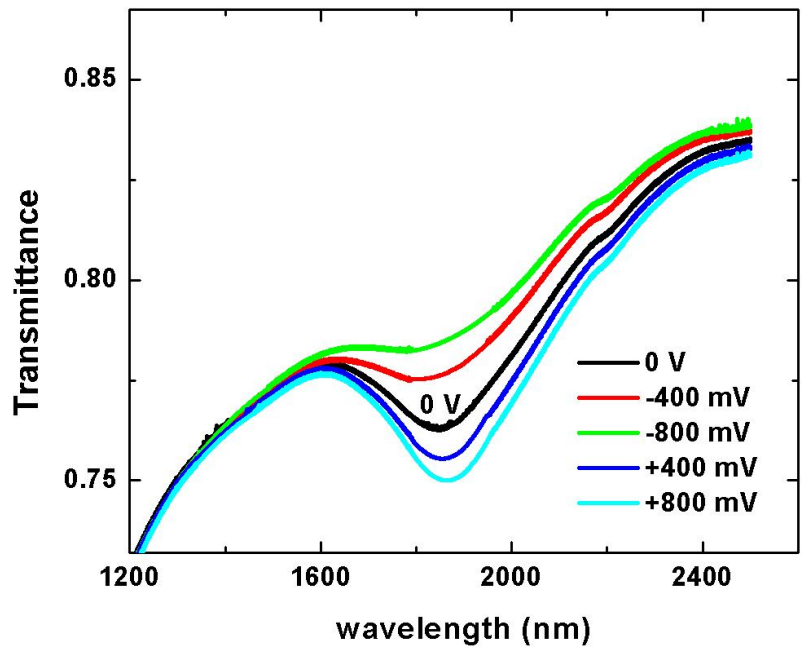

Figure 1. Spectral transmittance of the SWNT network as a function of applied gate voltage.

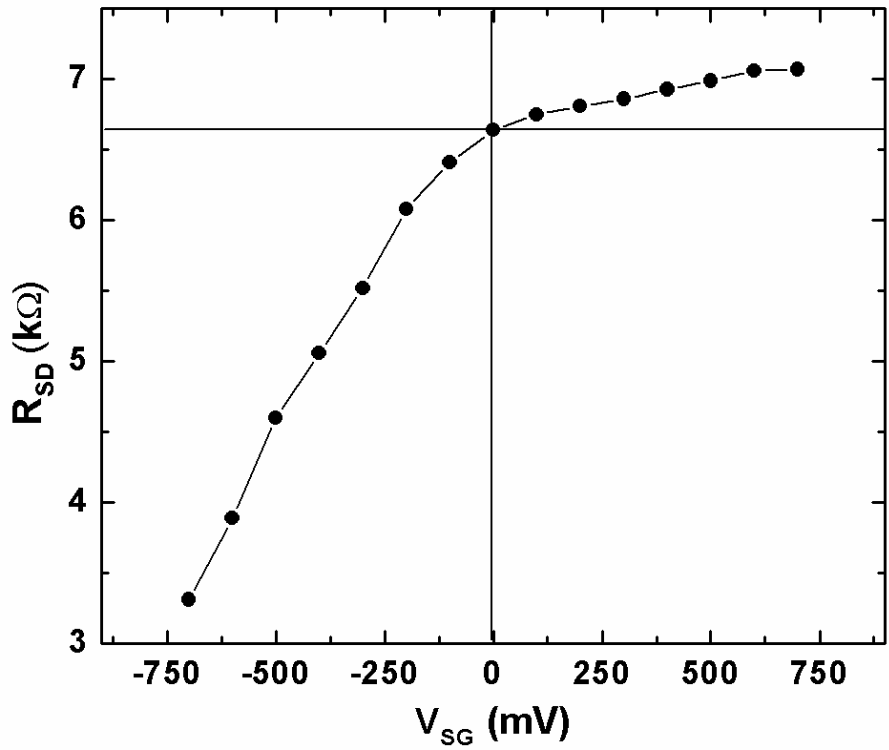

Figure 2. Network resistance vs. gate voltage.

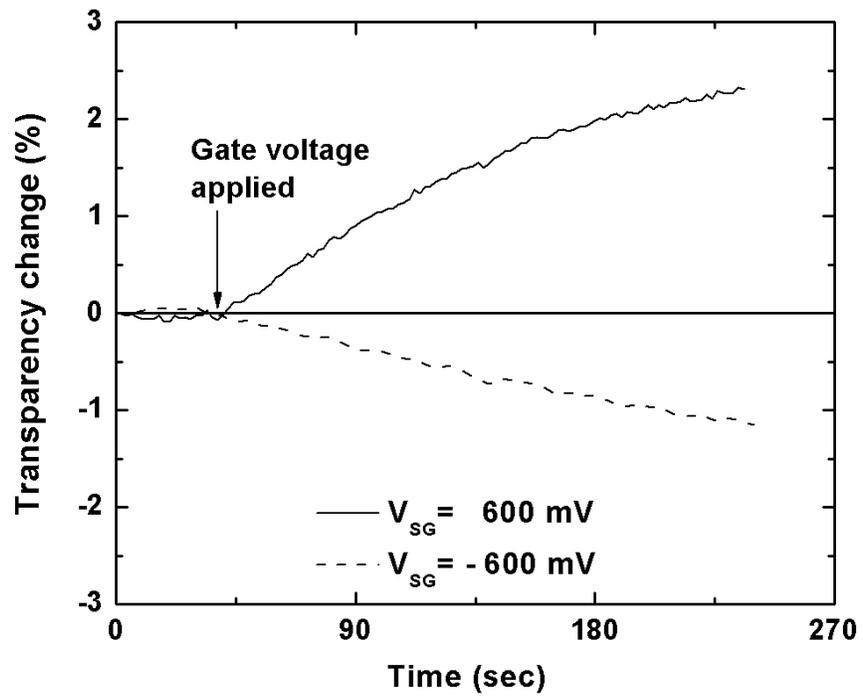

Figure 3. Transparency vs. time for positive and negative gate voltage bias.

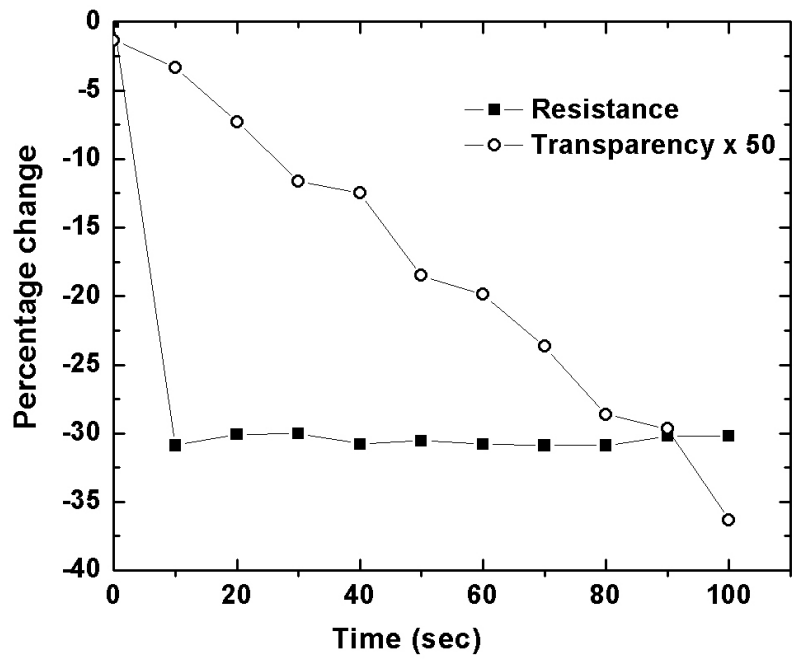

Figure 4. Rate of change for transparency (scaled 50 times) and network resistance (at $V_{SG}$ = -600 mV).

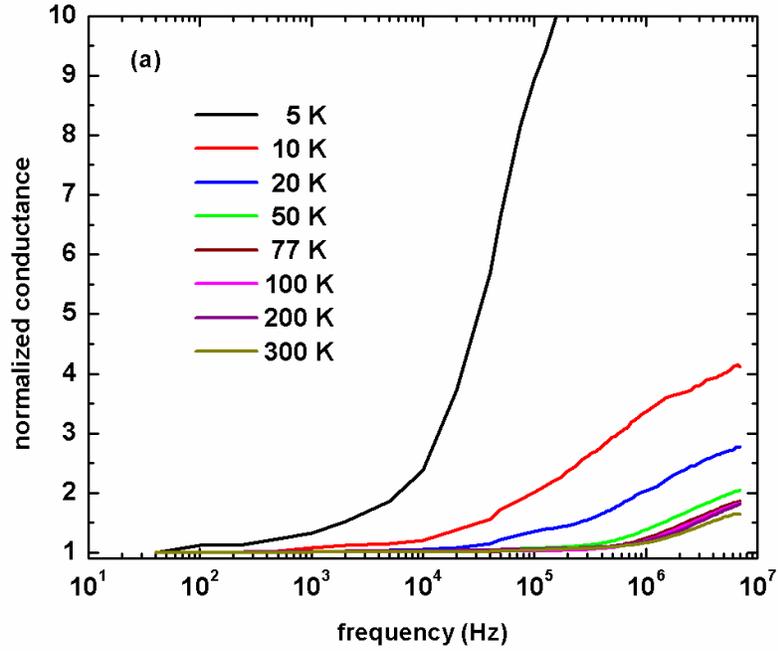

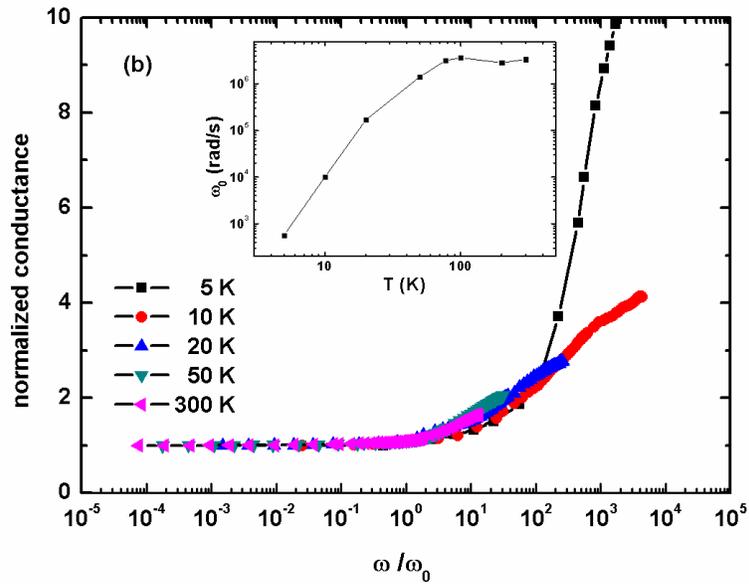

Figure 5. (a) Frequency dependence of normalized conductivity at various temperatures. (b) Scaling of the ac conductance at various temperatures [Inset: Onset frequency vs. temperature on a log-log scale].